# Gas Flow Rate Influence on Gas Temperature Regulation in a Reinforced Radio-Frequency Cross-Field Atmospheric Plasma Jet

Radhika T. P.[1], Satyananda Kar[1]

[1]*Plasma Applications Laboratory, Department of Energy Science and Engineering, Indian Institute of Technology Delhi, Hauz Khas, New Delhi 110016*

This study investigates the effect of gas flow rate on the gas temperature and discharge characteristics of a reinforced radio frequency cross-field atmospheric pressure plasma jet (APPJ) with an additional floating electrode. The plasma jet length, electron excitation temperature, electron density, and reactivity were enhanced by introducing copper floating electrodes of varying widths. However, this enhancement was accompanied by an undesired rise in gas temperature, limiting the plasma's application for heat-sensitive materials. To control this temperature rise, the gas flow rate varied from 1.5 to 9 lpm, showing a significant reduction in gas temperature from 438 K to 402 K as the flow rate increased, particularly at higher input powers. The study reveals that while an increase in gas flow rate initially improves ionization and reactivity by increasing electron excitation temperature and density, the insufficient input power for ionization at higher flow rates causes a decline in these parameters due to reduced ionization efficiency. Further optimization was achieved by increasing input power, which allowed better utilization of neutral atoms and improved plasma reactivity even at higher flow rates. The findings highlight the importance of tuning both gas flow rate and input power to maintain optimal plasma performance for various applications, particularly where controlled gas temperature and high reactivity are essential.

## 1 Introduction

Cold Atmospheric Pressure Plasmas (CAPs) have garnered significant interest due to their ability to generate reactive species at ambient conditions and as they are cost-effective compared to low-pressure systems with complex vacuum systems or magnetic fields for confinement [1,2, 3, 4]. Among the various plasma sources, radio frequency (RF) atmospheric pressure plasma jets (APPJs) are particularly notable for their efficient power coupling and lower breakdown voltages and ability to function as plasma chemical reactors, where reactions occur both within the discharge unit and in the surrounding air compared to other non-equilibrium plasma sources ignited by AC or pulsed DC [5, 6, 7]. The high concentration of energetic electrons in RF plasmas enhances the production of RONS, making these plasmas ideal for biomedical and material processing applications [8, 9]. In RF cross-field plasma jets, the high electric field between the electrodes creates a radially directed force that restricts the axial movement of electrons, leading to spatially confined ionization. This typically results in shorter and less reactive plasma plumes compared to linear-field jets [10]. To address the limitations of RF cross-field plasma jets, introducing an additional floating electrode has been shown to improve jet performance [11]. When this floating electrode is placed after the ground electrode, it acts as a secondary electron source, re-initiating the ionization wave that would otherwise slow down near the nozzle. This re-ignition of the ionization wave increases the plasma jet length, enhancing the interaction with ambient air and raising the ionization rate. As a result, this reinforced configuration of cross-field jet with an additional floating electrode is more suitable for applications requiring extended plasma plumes, such as treating biological targets and liquids that cannot tolerate vacuum environments. The presence of an additional floating electrode also increases electron density. It enhances plasma-air interactions, leading to a higher optical emission intensity of key reactive species such as OH, $N_2$, and O. This increase is directly linked to the plasma's reactivity, defined as its capability to generate and sustain chemically reactive species, including radicals (e.g., OH, O, and N), ions, and excited states. The improvement in plasma jet length and reactivity continues with increasing floating electrode width up to a certain threshold. Beyond this limit, however, excessive gas heating due to wider floating electrodes can lead to thermal instability, reducing the density of reactive species such as OH. Elevated temperatures not only destabilize the plasma, shrinking the active region for reactive species generation, but also accelerate recombination reactions, such as OH radicals forming stable molecules like water, thereby diminishing OH density. Since gas temperature plays a pivotal role in OH production, maintaining it within an optimal range is critical, particularly for applications involving heat-sensitive materials [12].

Since gas temperature rises with increasing floating electrode width, monitoring and controlling the temperature to optimize plasma reactivity without compromising material integrity becomes necessary. Gas flow rate is a crucial parameter that significantly influences gas temperature in plasma systems. Several studies have highlighted the influence of gas flow rate on plasma characteristics and reactive species generation in atmospheric pressure plasma jets (APPJs). Baek et al. demonstrated that varying the gas flow rate significantly impacts the generation of OH radicals during plasma-liquid interactions, thereby influencing plasma-induced chemical processes [13]. Yan et al.

conducted a computational study revealing that gas flow velocity affects the mixing of the working gas with ambient air, altering plasma dynamics such as axial and radial discharge structures, ion flux distribution, and the plasma interaction footprint on the substrate [14]. Experimental investigations by Li et al. [15] and Jin et al. [16] showed that the length of the luminous plasma jet in free space increases with gas flow rate under laminar conditions but decreases at higher flow rates due to turbulence-induced instability. Additionally, other researchers reported that higher gas flow rates lead to reductions in gas temperature and electron density within APPJs, emphasizing the complex interplay between flow dynamics and plasma behavior [17, 18, 19]. These findings collectively underscore the critical role of gas flow rate in shaping APPJ performance and its applicability to diverse processes. In this study, a cross-field radio frequency atmospheric pressure plasma jet, reinforced by adding an additional floating electrode, is characterized, aiming to create a plasma environment optimal for the treatment of thermally sensitive and biomedical materials (300 – 330 K). The impact of varying gas flow rates on the plasma discharge properties is thoroughly investigated. Key physical properties of the plasma plume, including electron excitation temperature, electron density, gas temperature, and species composition, are analyzed through optical emission spectroscopy, providing insights into the underlying mechanisms governing plasma behavior.

## 2 Experimental Setup

A reinforced cross-field plasma jet with an additional floating electrode is employed in this study. For a detailed schematic of the jet design and experimental setup, refer to ref. [11]. The plasma jet was generated inside a Pyrex glass tube with an inner diameter of 6 mm and a wall thickness of 2 mm. A 1.6 mm diameter copper power electrode was positioned at the centre of the tube, while a 3 mm wide copper strip wrapped around the tube served as the ground electrode. An additional floating electrode, in the form of a copper strip of different widths (3, 5, 8, and 10 mm), was placed after the ground electrode. A 13.56 MHz RF power supply, connected via a matching network (AIT-

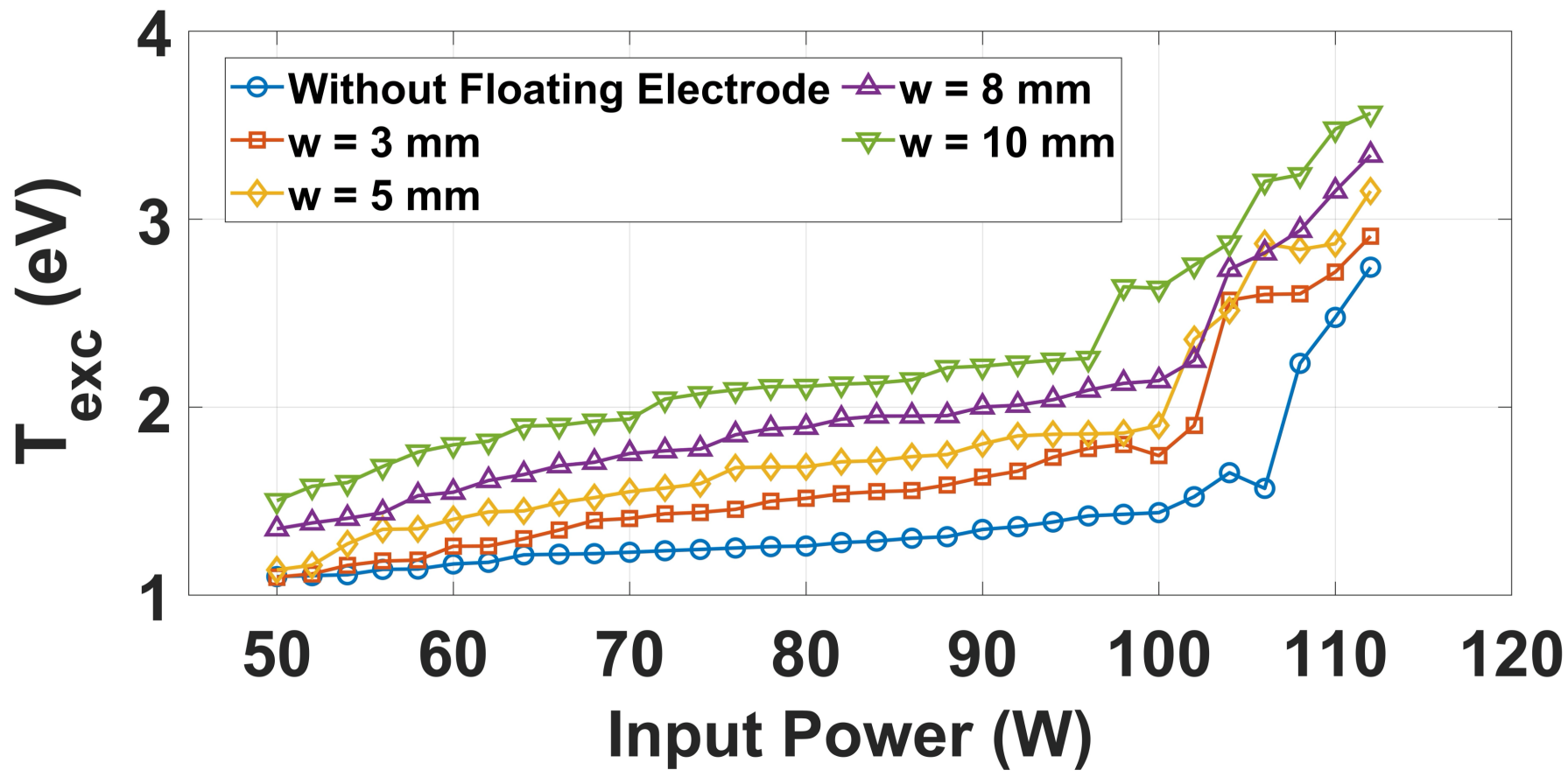

***Figure 1**: Variation in electron excitation temperature ($T_{exc}$) with input power for the jet with different widths of the additional floating electrode at a gas flow rate of 3 lpm.*

600) to ensure a 50 Ω load impedance, initiated the Argon discharge into the ambient air as a plasma jet. Voltage and current were measured using a calibrated Tektronix P6015A high-voltage probe and a Pearson Electronics 8590C current transformer, with actual power calculated from waveforms recorded on a KEYSIGHT DSOX3024T oscilloscope [20].

Gas temperature ($T_{gas}$) was measured using a calibrated K-type thermocouple (±2°C accuracy) with its open terminal enclosed in a thin glass capillary for insulation, ensuring minimal perturbation to the plasma plume. The thermocouple was floating and positioned at the tip of the plasma plume to capture the temperature without interference accurately. Optical emission spectroscopy (Ocean Optics HR4000) was employed to study the plasma jet's emissivity within the 200–900 nm wavelength range using a 200 μm optical fiber. The electron excitation temperature ($T_{exc}$) was determined through the Boltzmann plot method, while the electron density ($n_e$) was calculated using the line intensity ratio method by analyzing Ar I and Ar II spectral lines. For detailed methodology on the line ratio method, refer to Ref. [21].

In this study, the Argon gas flow rate was varied from 1.5 to 9 lpm to examine its effects on plasma parameters in jet equipped with an additional floating electrode of different widths (3, 5, 8, and 10 mm). Parameters such as plasma jet length, electron excitation temperature, electron density, plasma potential fluctuations, gas temperature, and the emission intensity of reactive oxygen and nitrogen species were analyzed.

## 3 Result and Discussion

In the plasma jet system with a cross-field electrode arrangement, plasma forms between the electrodes once the applied voltage exceeds the breakdown threshold, and as the input power increases, the jet extends outward into the ambient air. Adding the copper floating electrode, placed beneath the ground electrode, enhances both the plasma jet length and reactivity. This floating electrode enables the jet to exit the nozzle at lower power levels, with the minimum power needed for jet propagation decreasing as the floating electrode width increases. This demonstrates that wider floating electrodes extend the jet length at a given power. By confining electrons near the nozzle, the floating electrode creates an electron-rich region that serves as a source for the ionization wave, facilitating jet propagation with reduced power. As the width of the floating electrode increases, more electrons are concentrated in the centre of the glass tube, where the majority of gas flow occurs, leading to a longer plasma jet. The increased electron density enhances ionization, improving the plasma plume's length and reactivity [11]. The increment in electron excitation temperature $T_{exc}$ (Figure 1) and electron density $n_e$ (Figure 2) confirm the enhanced ionization as the electrode width grows. A notable increase in these parameters is observed beyond approximately 100 W, indicating a transition in the plasma regime. At lower input powers, the plasma operates in a stable glow discharge mode, characterized by moderate ionization and excitation rates. However, as the power increases beyond approximately 100 W, the enhanced electric field drives a higher rate of electron-neutral collisions, significantly increasing both ionization and excitation rates. The two distinct slopes observed in Figures 1 and 2 correspond to these operational regimes: the initial linear increase at lower powers represents the stable regime, while the steeper slope beyond 100 W indicates the transition to a more energetic regime.

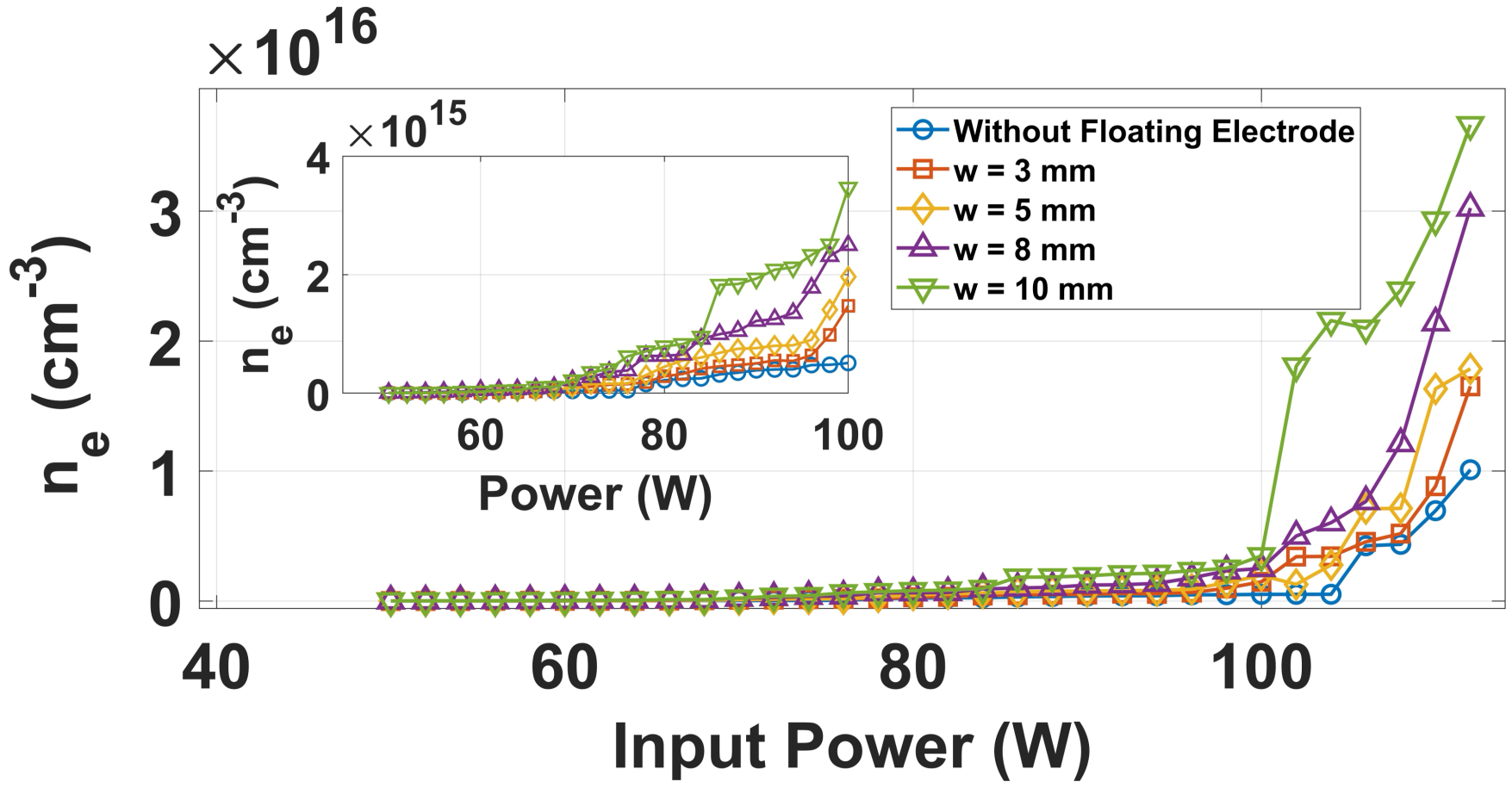

***Figure 2**: Variation in electron density ($n_e$) with input power for the jet with different widths of the additional floating electrode at a gas flow rate of 3 lpm.*

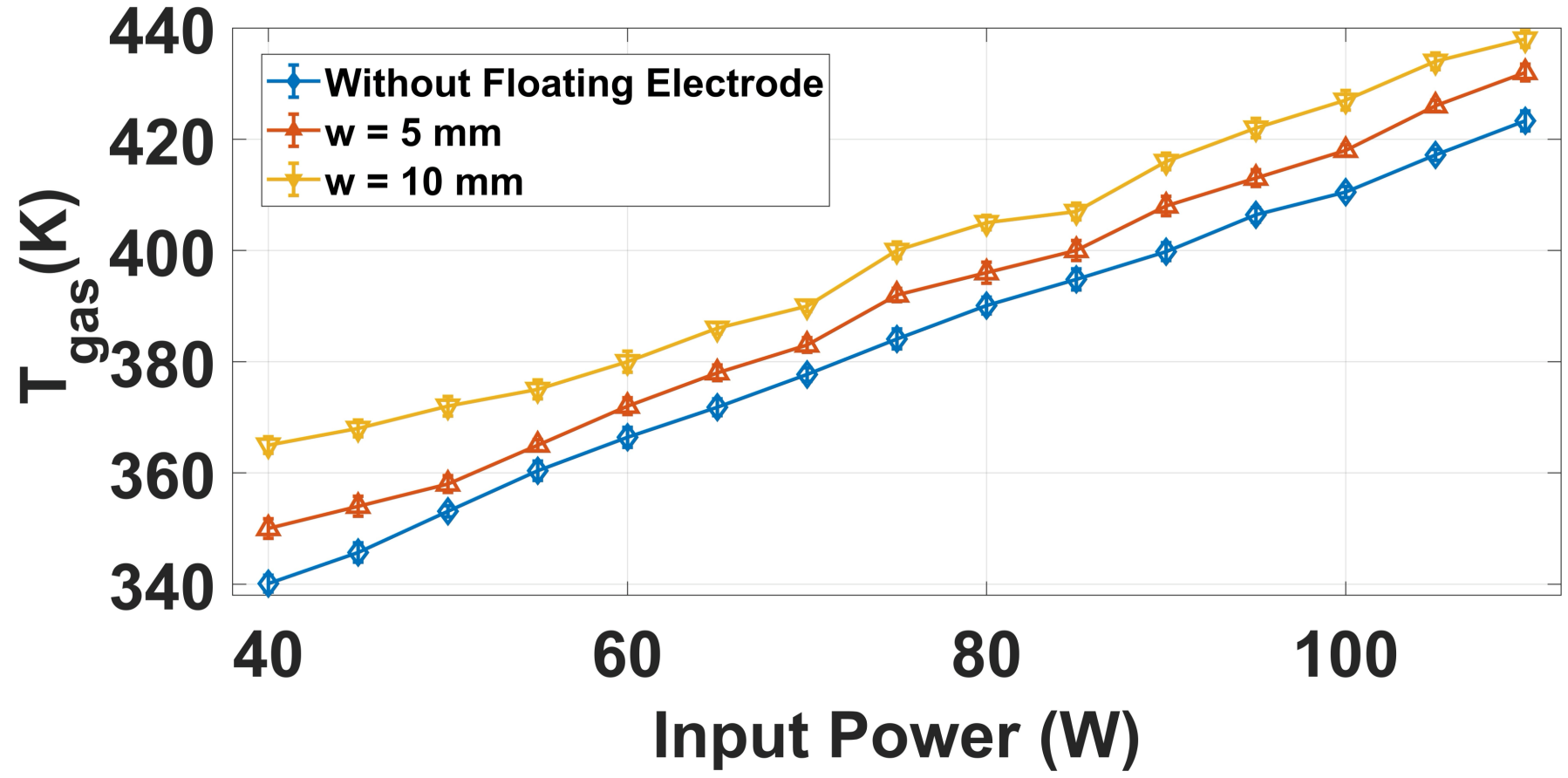

***Figure 3:** Variation in gas temperature ($T_{gas}$) with input power for different widths of the additional floating electrode at a gas flow rate of 3 lpm.*

While introducing an additional floating electrode significantly enhanced plasma jet length and reactivity, it also resulted in a notable rise in gas temperature, which poses challenges for applications involving heat-sensitive materials. Gas temperature measurements, taken with an insulated K-type thermocouple in the presence of the floating electrode of varying widths, revealed a steady increase in temperature with both increasing input power and electrode width (Figure 3). Gas temperature was raised from 423 K to 438 K for an input power of 110 W when the floating electrode width increased from 3mm to 10 mm. This uncontrolled temperature rise can lead to plasma instabilities, which in turn may cause plasma contraction. The observed increase in gas temperature with input power can be primarily attributed to Joule heating within the discharge region. As input power increases, the electric field strength intensifies, leading to a higher frequency of electron-neutral collisions. These collisions transfer energy from the electrons to neutral gas atoms, thereby increasing their kinetic energy and raising the gas temperature. Furthermore, the increase in gas temperature with the width of the floating electrode is associated with enhanced ionization near the electrode. This enhancement arises from the re-initiation of the ionization wave at the floating electrode due to the electric field generated by the accumulated charges on its surface. The intensified local field drives greater ionization activity, resulting in increased energy transfer to the gas atoms and, consequently, an elevated gas temperature. The elevated gas temperature impacts several key factors, including the interaction of the plasma plume with ambient air and target surfaces, as well as the concentration of reactive oxygen and nitrogen species (RONS). These factors are critical in plasma processing, especially in biomedical and material applications where precise control over reactive species is essential. Moreover, gas temperature plays a pivotal role in the production of OH radicals, which are vital for many plasma-induced chemical reactions. As the gas temperature increases, the relative emission intensity of OH radicals is affected, potentially leading to a decline in OH density due to thermal instabilities. The unchecked rise in gas temperature may trigger these instabilities, compromising both the plasma's stability and its effectiveness in generating crucial reactive species such as OH. This highlights the importance of carefully balancing gas temperature to maintain plasma performance and ensure suitability for heat-sensitive applications.

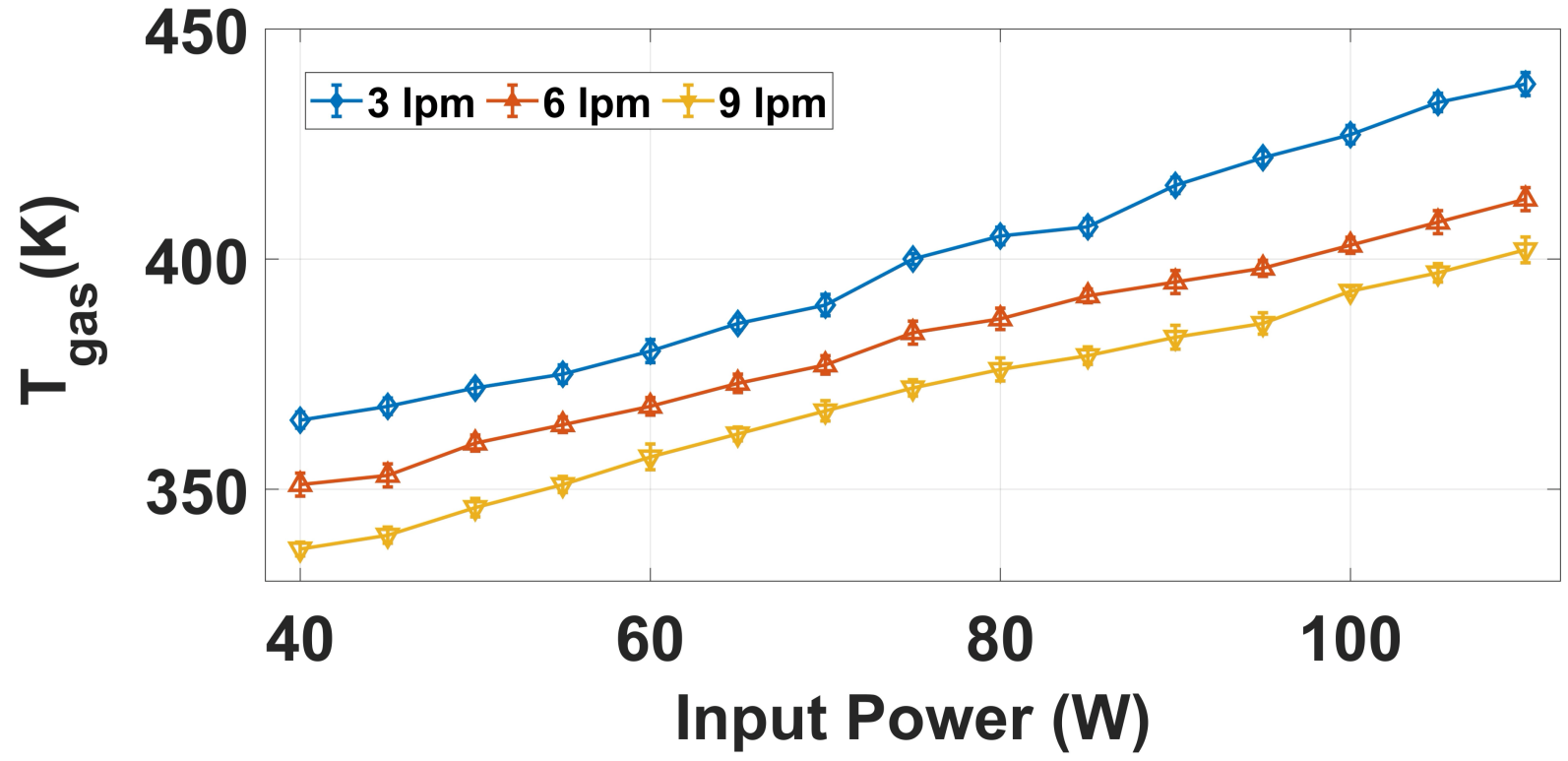

***Figure 4***: *Variation in gas temperature ($T_{gas}$) with input power for different gas flow rates for the jet with an additional floating electrode of width 10 mm.*

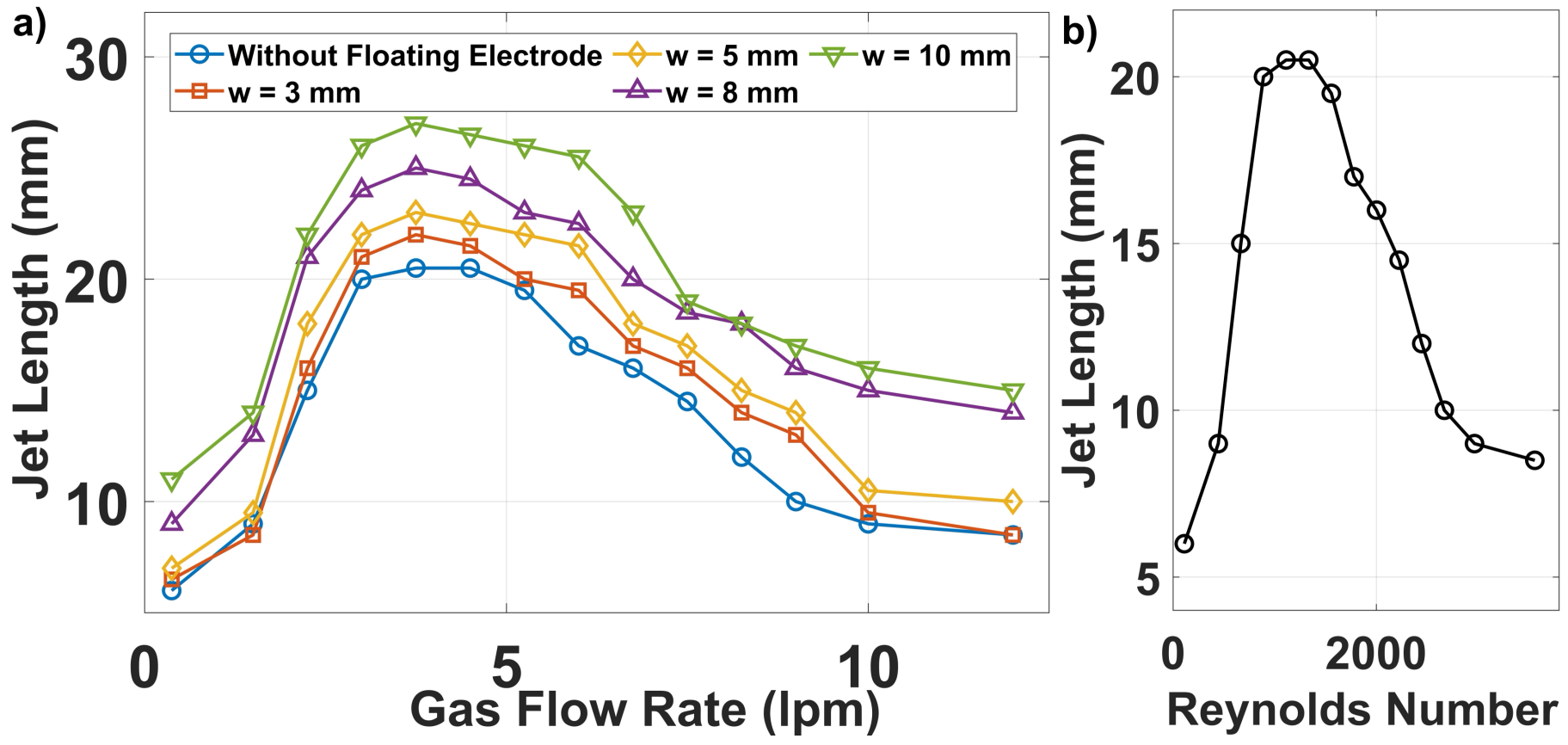

***Figure 5***: *a) Variation in plasma jet length with gas flow rate for the cross-field plasma jet with different widths of the additional floating electrode. b)Variation in plasma jet length with Reynolds number for the cross-field plasma jet without an additional floating electrode.*

Gas flow rate plays a critical role in dissipating the Joule heating generated during plasma discharge, thereby helping to reduce gas temperature [22]. To investigate the impact of gas flow rate on gas temperature, experiments were conducted with a reinforced cross-field plasma jet featuring a 10 mm wide additional floating electrode. The gas flow rate was varied from 3 lpm to 9 lpm. As shown in Figure 4, a significant reduction in gas temperature was observed as the flow rate increased. Specifically, at an input power of 110 W, the gas temperature decreased from 438 K to 402 K as the flow rate increased from 3 to 9 lpm. This observation highlights the importance of gas flow in carrying away the excess Joule heat produced within the discharge, which tends to increase gas temperature. By increasing the flow rate, the heated gas is effectively removed from the discharge region, preventing the accumulation of heat and mitigating the risk of thermal instability. The results clearly indicate that gas flow can be a valuable tool for controlling gas temperature in plasma systems.

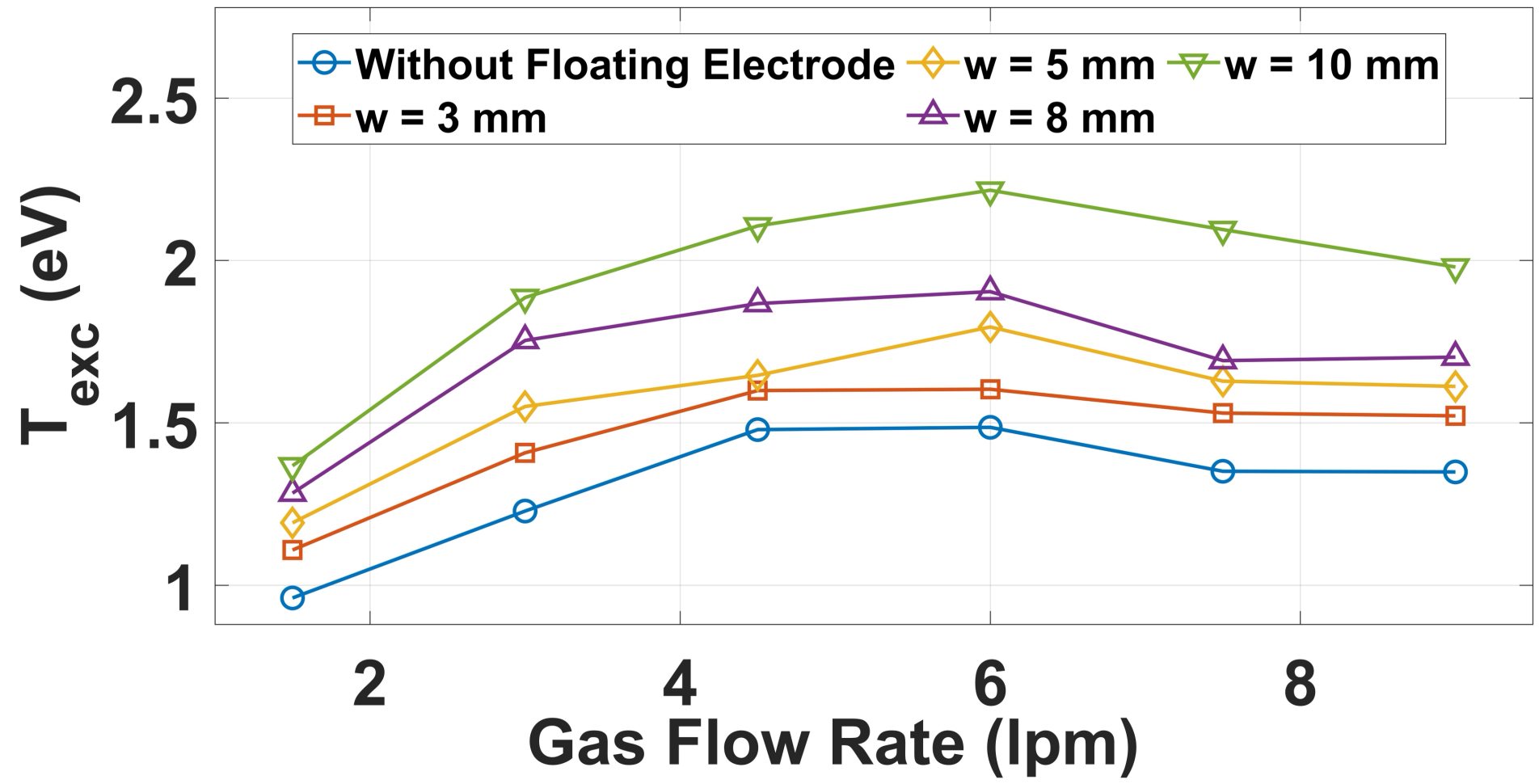

***Figure 6***: *Variation in electron excitation temperature ($T_{exc}$) with gas flow rate for the jet with different widths of the additional floating electrode at an input power of 85 W.*

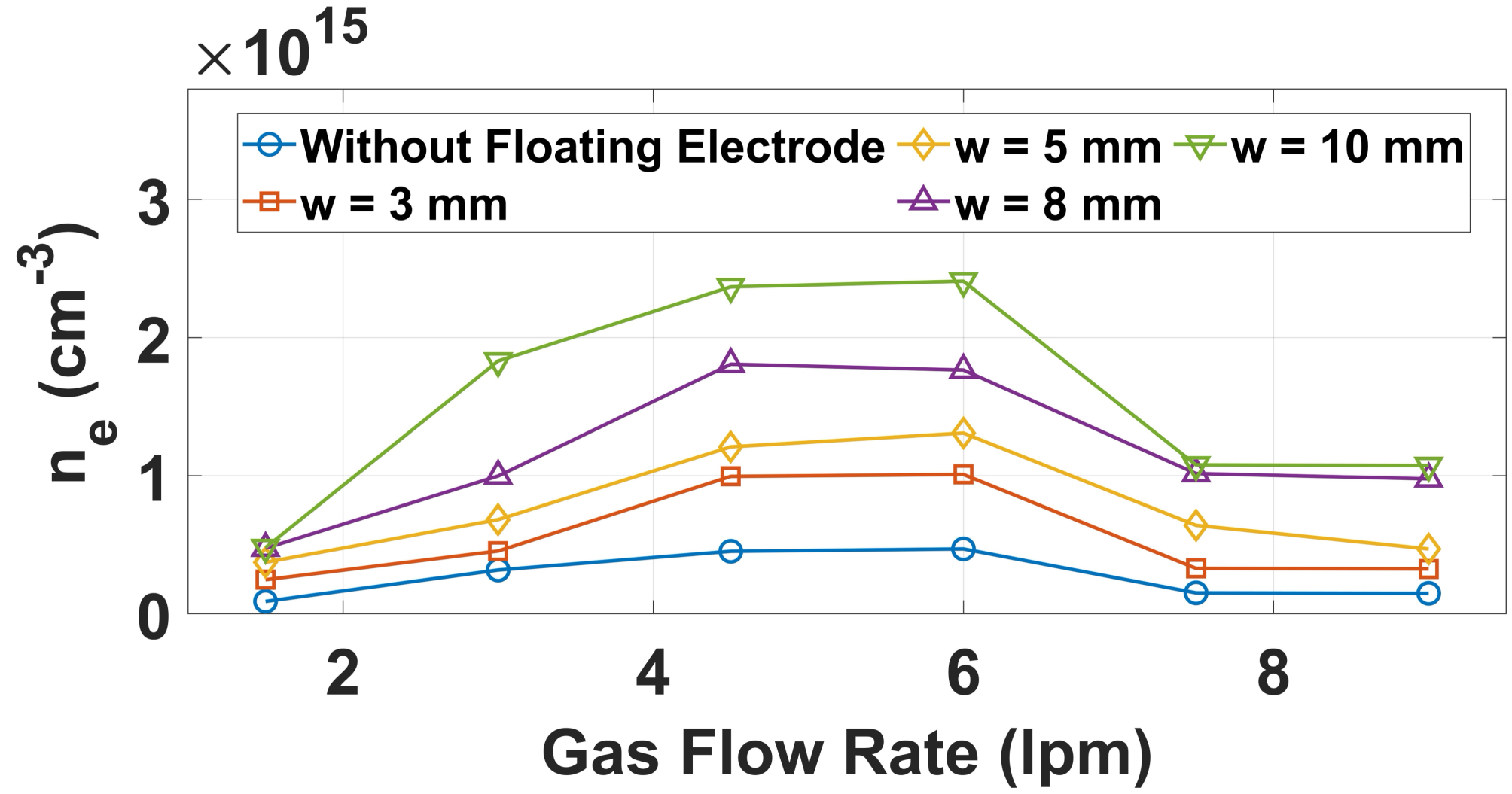

***Figure 7***: *Variation in electron density ($n_e$) with gas flow rate for the jet with different widths of the additional floating electrode at an input power of 85 W.*

Given that the gas flow rate effectively controls the gas temperature, it becomes essential to also examine how variations in flow rate influence other critical plasma parameters. These include plasma jet length electron excitation temperature ($T_{exc}$), and electron density ($n_e$), all of which are key indicators of the plasma's reactivity. A comprehensive understanding of how increasing gas flow rate affects these plasma parameters will provide deeper insights into optimizing the plasma jet for different applications, ensuring both its stability and effectiveness across a wide range of operational conditions. To begin the investigation, we analyzed the variation in plasma jet length with respect to the gas flow rate. The interaction rate between the plasma plume and the surrounding atmosphere is a key factor for most applications, as a higher interaction rate typically enhances the reactivity of the plasma plume. Many plasma-based processes, particularly in surface treatment and biomedical applications, require maximum plasma-air interaction to optimize the delivery of reactive species. As illustrated in Figure 5a, the plasma jet length initially increases with the gas flow rate, reaching a peak before gradually decreasing. This reduction in jet length at higher flow rates can be attributed to the transition from laminar to turbulent flow. The gas flow in the laminar flow regime is smooth and orderly, promoting stable plasma propagation and extended jet length. However, as the gas flow rate continues to rise, the flow becomes turbulent, leading to increased mixing with the surrounding air and a consequent reduction in jet length. This behavior was confirmed by calculating the Reynolds number for various gas flow rates, quantifying the transition from laminar to turbulent flow. The relationship between jet length and Reynolds number for the plasma jet without additional floating electrodes is provided in Figure 5b, showing a clear correlation between increasing Reynolds number and the onset of turbulence. The results highlight that while increasing the gas flow rate enhances plasma reactivity by increasing the jet length, excessive flow rates can disrupt the plasma jet's stability due to turbulence, limiting its effective length and reducing its interaction with the surrounding atmosphere. This balance between laminar flow stability and turbulence-induced shortening of the jet is crucial for optimizing plasma performance across different applications.

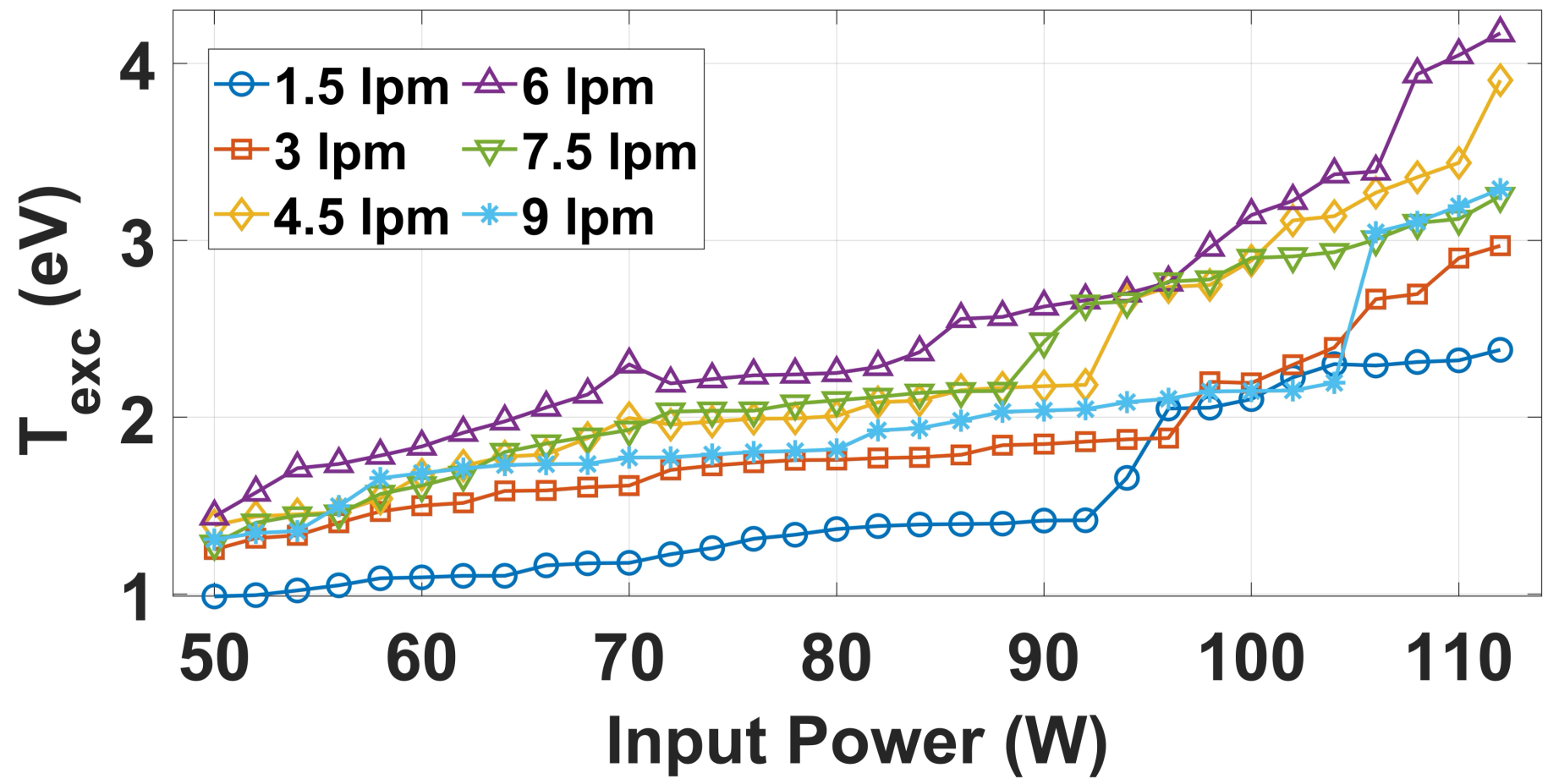

***Figure 8***: *Variation in electron excitation temperature ($T_{exc}$) with input power for different gas flow rates for the jet with an additional floating electrode of width 10 mm.*

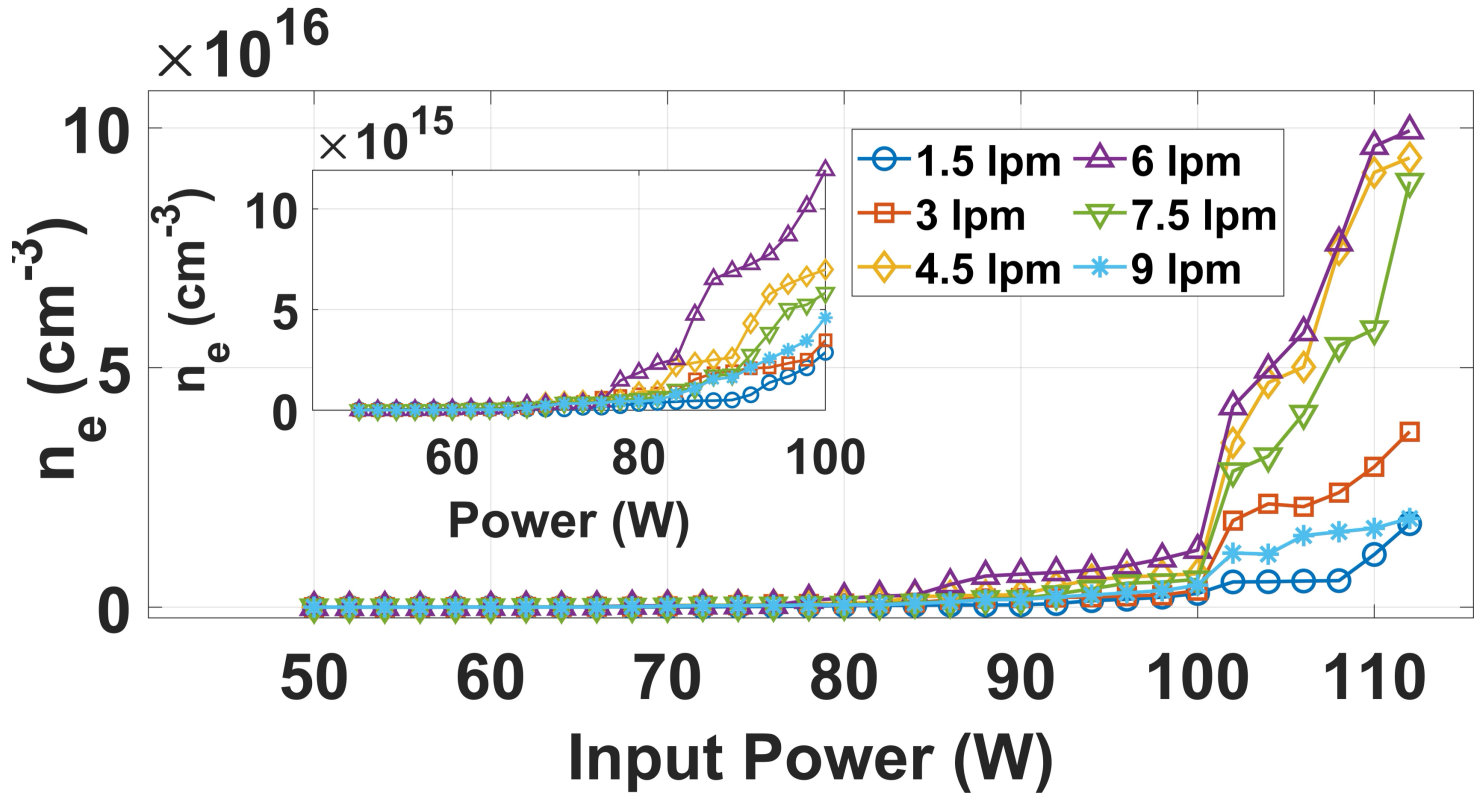

***Figure 9***: *Variation in electron density ($n_e$) with input power for different gas flow rates for the jet with an additional floating electrode of width 10 mm.*

The emission characteristics of the plasma jet at different gas flow rates were recorded to investigate the effect of flow rate on plasma parameters. Time-integrated emission spectra of the plasma plume were captured at a fixed position, located 10 mm axially and 4 mm radially from the jet nozzle. The recorded spectra revealed distinct emission lines corresponding to excited and ionized states of oxygen (O), nitrogen (N), and hydroxyl radicals (OH), along with prominent Ar I and Ar II lines. These emissions indicate the presence of various reactive species, which are critical for plasma reactivity and interaction with target surfaces in various applications. Analysis of the electron excitation temperature ($T_{exc}$) was conducted using the Boltzmann plot method, while electron density ($n_e$) was derived using the line ratio method by carefully selecting the appropriate Ar I and Ar II spectral lines. The results, depicted in Figures 6 and 7, show that both electron excitation temperature and electron density increase as the gas flow rate rises up to a certain threshold, after which both parameters begin to decline. This behavior is indicative of the complex interaction between gas flow and plasma dynamics. At a fixed input power, lower gas flow rates facilitate the supply of fresh argon into the discharge region, which enhances the interaction between electrons and neutral gas atoms. This increase in electron-neutral collisions promotes ionization, leading to a rise in both electron excitation temperature and electron density. As more neutral atoms are ionized, the plasma becomes more energized, contributing to improved ionization efficiency and overall plasma reactivity. The observed decrease in gas temperature ($T_{gas}$) with increasing flow rate, despite the rise in electron density and excitation temperature, can be explained by the interplay of gas flow dynamics and heat dissipation within the plasma jet. At higher flow rates, the introduction of fresh, cooler argon gas into the discharge region enhances convective cooling. This cooling effect outweighs the heating caused by electron-neutral collisions, resulting in an overall reduction in the gas temperature. However, as the gas flow rate continues to increase beyond a certain threshold, the fixed input power becomes insufficient to sustain the ionization of the larger volume of neutral atoms entering the plasma. The limited power cannot fully ionize the growing number of gas atoms, resulting in a saturation of ionization processes. In addition, the transition from laminar to turbulent flow at higher flow rates further destabilizes the plasma. Turbulence disrupts the orderly flow of gas, interfering with the plasma's energy distribution and reducing the efficiency of energy transfer between electrons and neutral atoms. This combination of insufficient power for ionization and turbulence-induced instability leads to a decline in both electron excitation temperature and electron density at higher flow rates. The increased mixing with ambient air cools the plasma, reducing the energy available for sustaining ionization processes and decreasing the overall plasma density. As a result, plasma performance diminishes, indicating that there is an optimal gas flow rate range where ionization and plasma stability are maximized for a given input power.

To counteract the reduced ionization caused by increased gas flow rates, which are often adjusted to manage gas temperature, one effective solution is to raise the input power. Higher input power supplies the necessary energy to ionize the larger volume of neutral atoms introduced by the elevated flow rates, thereby enhancing the plasma's overall reactivity. This approach ensures that even at higher flow rates, sufficient ionization occurs, leading to improved plasma performance. As illustrated in Figures 8 and 9, both electron excitation temperature and electron density rise with increasing input power. This trend confirms that boosting the input power compensates for the reduced ionization efficiency typically associated with higher gas flow rates. By increasing input power, more neutral atoms are ionized, leading to a more energetic and reactive plasma plume, which can better meet the demands of various applications. Understanding the relationship between gas flow rate, input power, and plasma parameters is essential for optimizing the reinforced plasma jet system with an additional floating electrode. By carefully tuning these parameters, it is possible to strike a balance between maintaining a low gas temperature—important for heat-sensitive applications—and achieving high plasma reactivity. This flexibility allows for the customization of the plasma jet to specific application requirements, ensuring that the system delivers controlled gas temperatures while maintaining the necessary ionization and reactivity for effective plasma treatments.

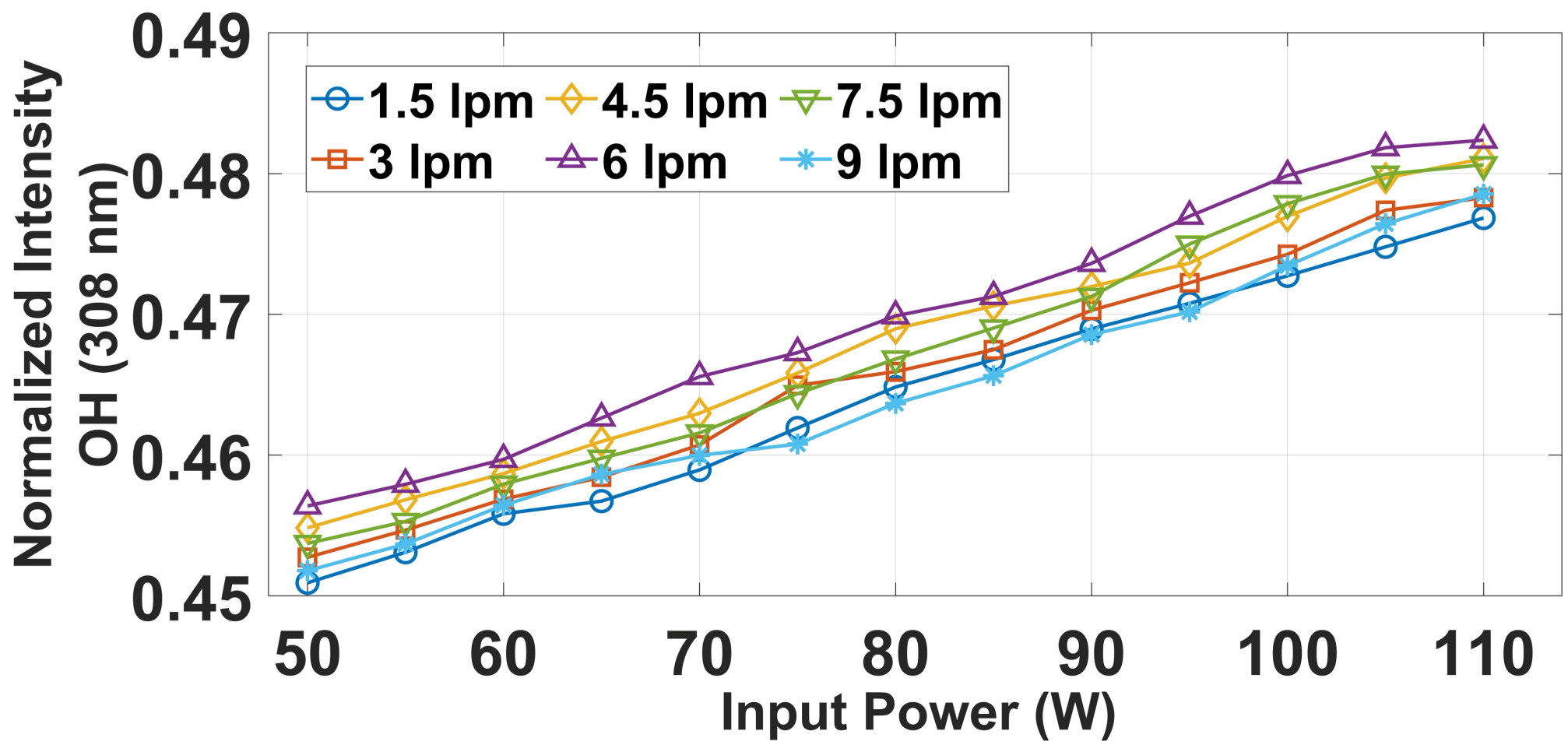

***Figure 10***: *Variation in normalized optical emission intensity of OH with input power for different gas flow rates for the jet with an additional floating electrode of width 10 mm.*

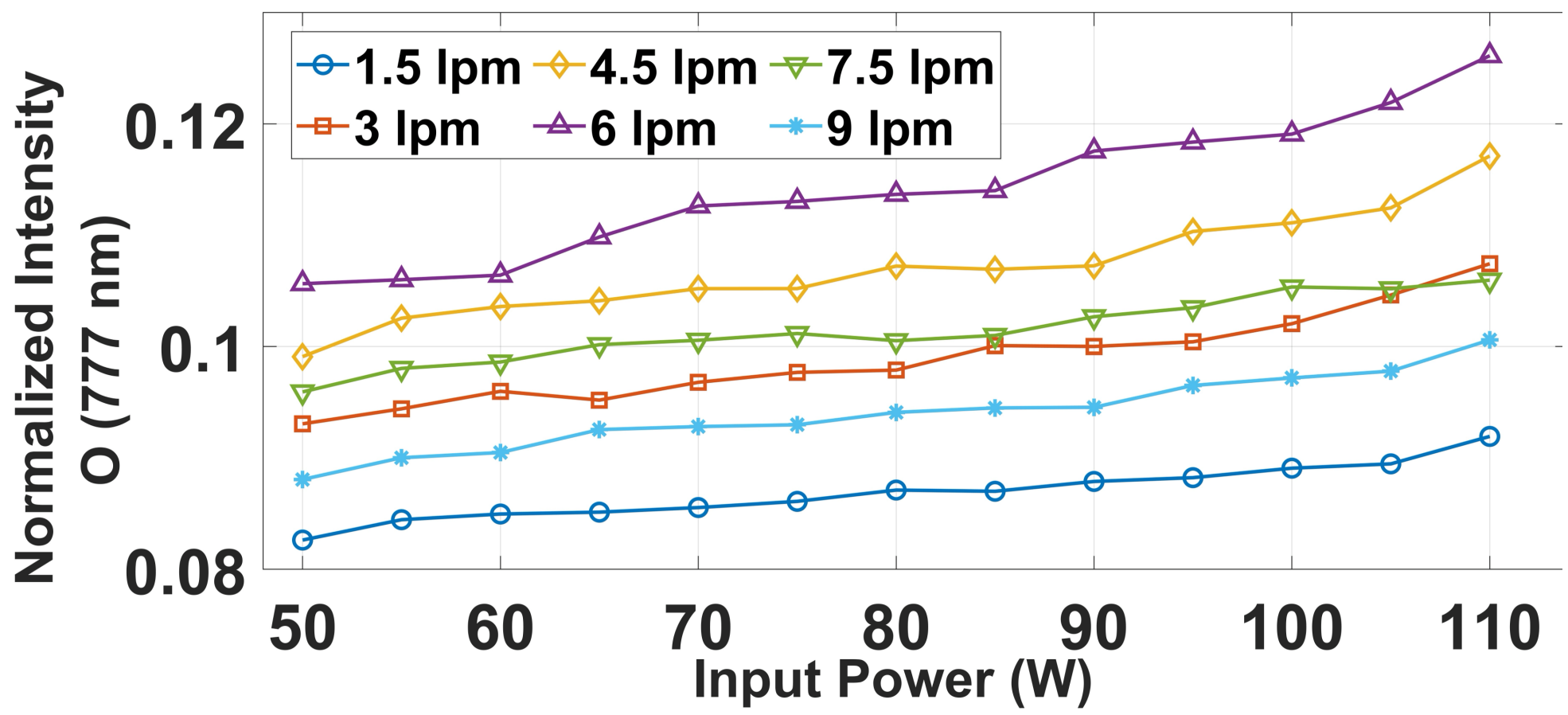

***Figure 11****: Variation in normalized optical emission intensity of O with input power for different gas flow rates for the jet with an additional floating electrode of width 10 mm.*

Building on our efforts to manage the gas temperature in a reinforced cross-field plasma jet with an additional floating electrode while improving reactivity, the relative optical emission intensities of reactive oxygen and nitrogen species (RONS) provide critical insights into the plasma behavior. Figures 10, 11, and 12 depict the emission intensities of key species such as OH (308 nm), $N_2$ (337 nm), and O (777 nm), revealing a characteristic trend: the intensities increase with gas flow rate initially, then decrease beyond a certain point. This pattern is attributed to the dynamics of ionization, where enhanced gas flow introduces more neutral atoms into the plasma, promoting ionization at first, but as flow rates continue to rise, the ionization efficiency diminishes due to insufficient input power and turbulence, leading to reduced emission intensities. To counterbalance this reduction in ionization at higher flow rates, we implemented the strategy of increasing input power, which successfully restored the ionization levels. As demonstrated in the figures (Fig.s 10-12), the relative emission intensity of OH, O, and $N_2$, increases with input power, indicating a stronger ionization process. This finding underscores that more neutral atoms are ionized by elevating input power, enhancing the plasma reactivity even at higher flow rates. Given that the reinforced cross-field plasma jet with an additional floating electrode exhibits improved reactivity and jet length, controlling the gas temperature without sacrificing plasma performance becomes critical, especially for applications involving heat-sensitive materials. The results clearly show that by strategically tuning both gas flow rate and input power, it is possible to achieve the desired plasma parameters while maintaining a balance between temperature control and plasma reactivity.

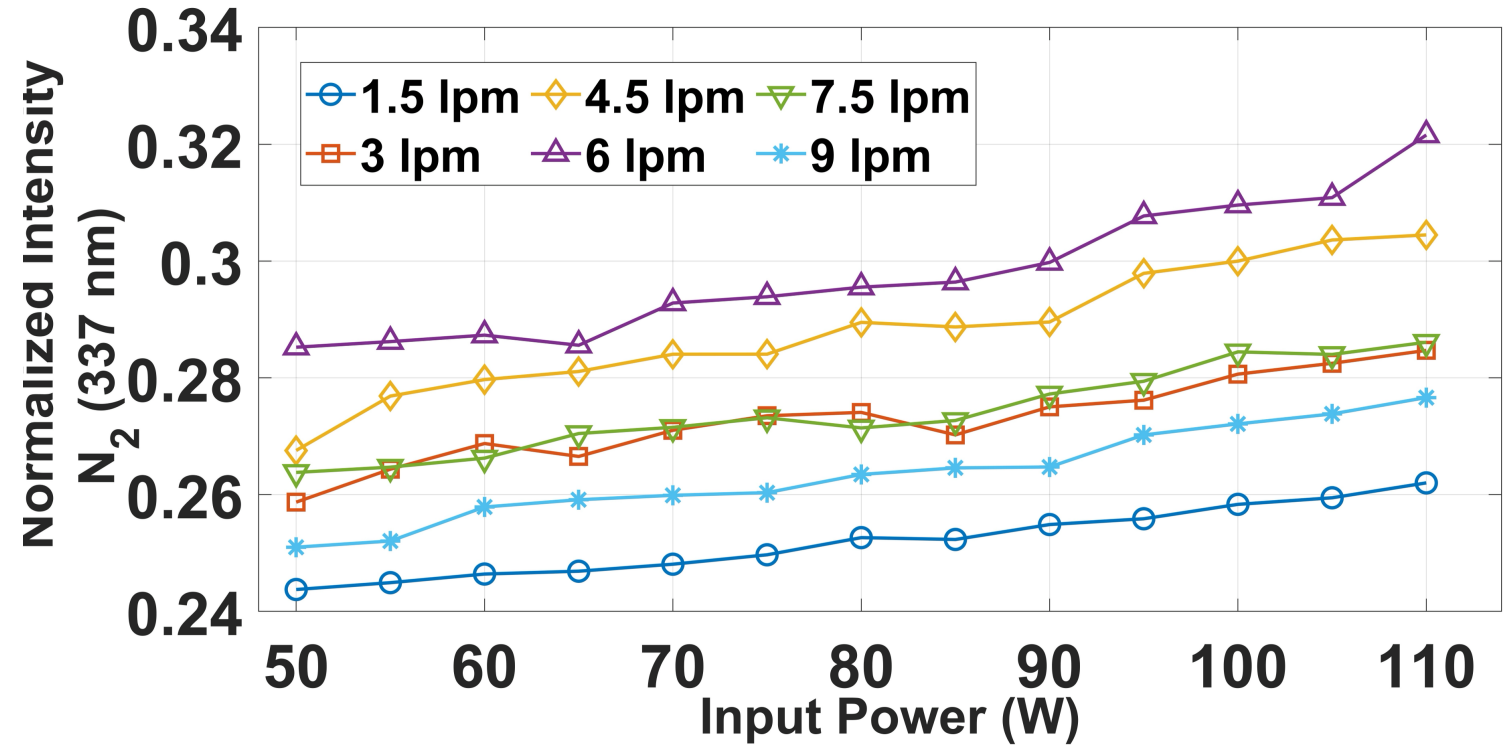

***Figure 12****: Variation in normalized optical emission intensity of $N_2$ with input power for different gas flow rates for the jet with an additional floating electrode of width 10 mm.*

## 4 Conclusion

This study explored the performance of a reinforced cross-field plasma jet with an additional floating electrode, focusing on controlling gas temperature and enhancing plasma reactivity. The introduction of floating electrodes significantly increased plasma jet length and reactivity by confining electrons and enhancing ionization. However, it also led to a rise in gas temperature, which poses challenges for heat-sensitive applications. To mitigate this, we investigated the effect of gas flow rate, which effectively reduced gas temperature by dissipating heat generated during plasma discharge. Initially, electron excitation temperature, electron density, and reactivity increased with flow rate due to the availability of more neutral particles, but beyond a certain point, they declined as the input power became insufficient to ionize the additional particles. To counter this, increasing input power restored plasma performance, ensuring sustained ionization and higher reactivity at elevated flow rates. The findings suggest that by strategically balancing gas flow rate and input power, the plasma jet system can be fine-tuned for specific applications, offering a flexible solution that combines enhanced reactivity with controlled gas temperature. The optimal region for the jet, however, is application-dependent and can be identified by targeting the desired range of plasma parameters such as electron density, excitation temperature, and RONS concentration, while maintaining an acceptable gas temperature. This adaptability allows for the precise selection of input power and flow rate to achieve the best match for each application, ensuring that the plasma jet's performance is optimized for various industrial and scientific uses.

## Acknowledgements

Radhika T.P. sincerely acknowledges the insightful contributions and support from Aishik Basu Mallick, Sarthak Das, Tejashwi Rana, Suryasunil Rath, and Pratyay Chattopadhyay during the course of this research.

*Corresponding author*
*e-mail:* satyananda@dese.iitd.ac.in (Satyananda Kar)